\documentclass[onecolumn,showpacs,preprintnumbers,amsmath,amssymb]{article}
\usepackage{xcolor}
\usepackage{amssymb,amsmath}
\usepackage{graphicx}
\usepackage{hyperref}
\usepackage[affil-it]{authblk}

\begin{document}

\title{Revisiting Unimodular Quantum Cosmology}

\author[1]{Júlio C. Fabris \footnote{E-mail: julio.fabris@cosmo-ufes.org}}
\author[1,2]{Luiz Filipe Guimarães}
\author[3]{Nelson Pinto Neto}
\author[4]{Mahamadou Hamani Daouda}

\affil[1]{Núcleo Cosmo-Ufes \& Departamento de Física, UFES, Vitória, ES, Brazil}
\affil[2]{{Departamento de F\'isica, Universidade Estadual de Londrina, Londrina, PR, Brazil}}
\affil[3]{{Centro Brasileiro de Pesquisas Físicas, Rio de Janeiro, RJ, Brazil}}
\affil[4]{{Département de Physique - Université de Niamey, Niamey, Niger}}

\maketitle

\begin{abstract}
The quantization of unimodular gravity in minisuperspace leads to a time evolution of states generated by the Hamiltonian, as in usual quantum mechanics. We revisit the analysis made in Ref. \cite{unruh}, extending it to phantom scalar fields. It is argued that only in this case a non-trivial evolution for the scalar field can be obtained. The behavior of the scale factor presents a bounce followed by a de Sitter expansion, reproducing the quantum cosmological scenario in General Relativity when the source is given by a cosmological term described by the Schutz variable. The analysis is extended to the Brans-Dicke scalar tensor theory.
\end{abstract}

\section{Introduction}

General Relativity is a theory invariant by time reparametrization, which is encoded in the lapse function that appears explicitly, for example, in the $1 + 3$ decomposition of the metric. The $1 + 3$ decomposition is based in foliating the four dimensional space-time by three-dimensional hyper-surfaces defined by constant time $t$. The line element can then be written as,
\begin{eqnarray}
ds^2 = g_{\mu\nu}dx^\mu dx^\nu = (N^2 - N^i N_j)dt^2 - 2 N_i dx^i dt - h_{ij}dx^i dx^j,
\end{eqnarray}
where $N$ is the lapse function and $N_i$ is the shift function. Both $N$ and $N_i$ describe the evolution of hyper-surfaces. The metric on the space-like hyper-surface is given by $h_{ij}$ and $N_i = h_{ij}N^j$. For a detailed exposition of the $1 + 3$ (also called Arnowitt–Deser–Misner decomposition, or ADM), see Ref. \cite{nelson} and references therein. This decomposition allows for a Hamiltonian formulation of General Relativity (GR) theory, in which $N$ appears as a Lagrange multiplier, implying that the resulting Hamiltonian must vanish. Under canonical quantization, this construction leads to a Schrödinger-like equation with no time variable. In brief, this is the problem of time of canonical quantum gravity. It is related to the constraints existing in the GR theory due to its invariance by general diffeomorphical transformations.

One of the problems of quantum gravity is to recover the notion of time and, consequently, to obtain the dynamical evolution of the system \cite{kuchar}. There are many different proposals to recover the time evolution. One possibility is to introduce matter fields containing dynamical variables. An example of such approach is given in Refs. \cite{ruba,demaret,nivaldo1,nivaldo2}.
%%%LF: repetido -- There are many other different proposals o how ton recover the time evolution. One of them is by introducing matter fields containing dynamical variables \cite{ruba,demaret,nivaldo1,nivaldo2}.
In these references, the matter fields are described by the Schutz formalism \cite{schutz1,schutz2} resulting in a linear conjugated momentum, related to the matter field, which is interpreted as the time variable.

Another possibility to recover the time evolution in Quantum cosmology (QC) is to break the time reparameterization invariance. This can be achieved, for example, in Unimodular Gravity (UG). In UG, a constraint is introduced in the Einstein-Hilbert action by fixing the determinant of the metric equal to one. This choice imposes a condition in the coordinate system to be used; in particular it fixes the lapse function.  At the same time, such a constraint leads to trace-free gravitational equations. The information concerning the Ricci scalar is somehow lost, a consequence of the unimodular condition which breaks the general diffeomorphism invariance of GR, reducing it to a transverse diffeomorphism, see Ref. \cite{trans}. UG can be extended in order to recover the freedom of choice of the time coordinate by imposing a fiducial metric, that acts as an external field, not subject to the variational principle at the action level.  The trace-free character of UG implies that the cosmological constant disappears from the dynamical equations. In order words, UG somehow {\it degravitates} the vacuum energy, classically interpreted as a cosmological constant. However, the classical cosmological constant remains somehow hidden in the structure of UG,  as it reappears as an integration constant. The cosmological constant problem is, at least, alleviated since, as an integration constant, the connection with the vacuum energy is lost.

In Ref. \cite{unruh} the construction of a quantum cosmological model has been discussed using UG in its original formulation, without a fiducial metric, in presence of a scalar field. It has been shown that, in the minisuperspace, a Schrödinger-type equation is obtained, with the Hamiltonian driving the time evolution of the system.  Since in this original formulation of UG, fixing $\sqrt{-g} = 1$ breaks the invariance by time reparameterization, the time coordinate is obtained similarly to the ordinary quantum mechanics. The resulting Schrödinger-type equation is solved and the stationary states are settled. It is shown that the universe reaches a de Sitter phase for large values of the scale factor. Hence, as it happens in the classical UG, the cosmological constant is somehow {\it hidden} in the UG formalism. In addition, this result has the nice feature that it is in qualitative agreement with the accelerated expansion of the universe observed today.

The present note revisits the problem of constructing a quantum cosmological model in UG, performing some extensions of the seminal work of Ref. \cite{unruh}. We will discuss in more details the Schrödinger-type equation, the question of the positiveness of the energy and the introduction of phantom fields besides the ordinary self-interacting scalar field. A wave packet is explicitly constructed, from which the expectation value for the scale factor and the scalar field is determined: a bounce universe, with asymptotic de Sitter phase, is obtained, while the expectation value of the scalar field indicates a constant value. The comparison with the Schutz approach for quantum cosmology in GR with a cosmological constant is made. The results are essentially the same in UG quantum cosmology and in GR quantum cosmology with a cosmological constant described through the Schutz formalism. The generalization to a Brans-Dicke UG model is also discussed, both and in the non-minimal and minimal coupling formulations.

In next section, Unimodular Gravity is briefly presented. In section 3, the Unruh approaches is described. In section 4, the positiveness of energy and the self-adjoint character of the Hamiltonian are discussed. Explicit solutions, in the absence of the potential term, are obtained in section 5. The wave packets are constructed in section 6 and the expectation values for the observables are computed as well. The Brans-Dicke theory is discussed in section 7. Final comments are presented in section 8.

\section{Unimodular gravity}

The Unimodular equations can be deduced by using the Einstein-Hilbert Lagrangian with an extra condition imposed on the determinant of the metric implemented, for example, by using Lagrange multipliers. On general grounds, the Lagrangian reads,
\begin{eqnarray}
\label{l1}
{\cal L} = \sqrt{-g}R + \xi(\sqrt{-g} - \chi) + {\cal L}_m.
\end{eqnarray}
where $\xi$ is a Lagrange multiplier, $\chi$ is a non-dynamical reference density tensor and ${\cal L}_m$ is the matter Lagrangian. This constrained Lagrangian leads to the UG field equations:
\begin{eqnarray}
R_{\mu\nu} - \frac{1}{4}g_{\mu\nu}R = 8\pi G\biggr\{T_{\mu\nu} - \frac{1}{4}g_{\mu\nu}T\biggl\}.
\end{eqnarray}
These equations are trace-free, hence no information on the Ricci scalar $R$ can be obtained.

Is it possible to obtain a quantum cosmological model of this structure?
Let us investigate this in the mini-superspace approach, that is, in the framework of a flat Friedmann-Lemaître-Robertson-Walker (FLRW) metric. The gravitational part of  Lagrangian (\ref{l1}) becomes (imposing $16\pi G = 1$),
\begin{eqnarray}
\label{L1}
{\cal L}_g = 6\frac{a\dot a^2}{N} + \xi(Na^3 - \chi).
\end{eqnarray}
From this expression, we obtain the conjugate momentum,
\begin{eqnarray}
\pi_a = 12\frac{a\dot a}{N}.
\end{eqnarray}
The gravitational Hamiltonian takes then the form,
\begin{eqnarray}
\label{H1}
{\cal H}_g = N\biggr\{\frac{1}{24}\frac{\pi_a^2}{a} + \xi a^3\biggl\} - \xi\chi.
\end{eqnarray}
We remark that the reference tensor density and the Lagrange multiplier appear explicitly in the Hamiltonian. The reference tensor density will not appear in the corresponding Schrödinger equation after solving the constraint.

Notice that the variation of (\ref{L1}) with respect to $\xi$ leads to,
\begin{eqnarray}
\chi = Na^3.
\end{eqnarray}
Inserting this result in (\ref{H1}) leads to the Hamiltonian,
\begin{eqnarray}
{\cal H}_g = \frac{N}{24}\frac{\pi_a^2}{a}.
\end{eqnarray}

In this general approach to Unimodular gravity, the usual Hamiltonian expression in the minisuperspace is recovered. In particular, the Hamiltonian constraint remains the same as in the GR theory:
\begin{eqnarray}
{\cal H} = N H, \quad H = \frac{1}{24}\frac{\pi_a^2}{a}.
\end{eqnarray}
This leads to $H = 0$, implying, after quantization, in a Schrödinger-type equation without an explicit time coordinate:
\begin{eqnarray}
H \Psi = 0.
\end{eqnarray}
The time reparameterization freedom of GR is encoded here in the choice of the external function $\chi$. However, in the the original Unimodular proposal $\chi = 1$, and the lapse function is fixed.

\section{Revisiting Unruh approach}

Unruh in Ref. \cite{unruh} follows a simpler approach. He chooses $\chi = 1$, leading to,
\begin{eqnarray}
N = \frac{1}{a^3}.
\end{eqnarray}
This choice fixes the lapse function, with the obvious consequence that the theory is no longer invariant by time reparameterization, and the Hamiltonian becomes the generator of the time evolution of states, as in the usual quantum mechanics.

The original action describing gravity coupled with a self-interacting scalar field is given by,
\begin{eqnarray}
S = - \int \biggr\{R - \frac{\epsilon}{2}\phi_{;\rho}\phi^{;\rho} + V(\phi)\biggl\}\sqrt{-g}d^4x.
\end{eqnarray}
In this expression, $\epsilon = \pm 1$, representing either an ordinary scalar field (positive) or a phantom scalar field (negative). In Ref. \cite{unruh} only the case $\epsilon = + 1$ has been considered. Here, we leave open both possibilities, that is, $\epsilon = \pm 1$.
Imposing a flat FLRW metric and after discarding a surface term, the action reduces to,
\begin{eqnarray}
S = - \int\biggr\{6\frac{a\dot a^2}{N}  - \frac{\epsilon}{2}\frac{a^3}{N}\dot\phi^2 + N a^3V(\phi)\biggl\}d^4 x.
\end{eqnarray}
Inserting the unimodular condition $Na^3 = 1$, the action becomes,
\begin{eqnarray}
S = - \int\biggr\{6a^4\dot a^2  - \frac{\epsilon}{2}a^6\dot\phi^2 + V(\phi)\biggl\}d^4 x.
\end{eqnarray}

The final Lagrangian is
\begin{eqnarray}
\label{lug}
{\cal L} =  - 6a^4\dot a^2  + \frac{\epsilon}{2}a^6\dot\phi^2 - V(\phi).
\end{eqnarray}
The conjugate momenta are,
\begin{eqnarray}
\pi_a &=& - 12a^4\dot a,\\
\pi_\phi &=& \epsilon a^6\dot \phi.
\end{eqnarray}
Using the usual Legendre transformation, the Hamiltonian reads \footnote{Note that in Ref. \cite{unruh} the first term has $a^6$ in the denominator: it is clearly a misprint.},
\begin{eqnarray}
\label{ha}
{\cal H} = - \frac{1}{24}\frac{\pi^2_a}{a^4} + \frac{\epsilon}{2}\frac{\pi^2_\phi}{a^6} + V.
\end{eqnarray}
%Note that in Ref. \cite{unruh} the first term has $a^6$ in the denominator: it is clearly a misprint.

Introducing the usual quantization procedure with a particular factor ordering, and defining $z = a^3$, the Schrödinger-like equation becomes,
\begin{eqnarray}
\label{ea}
 \frac{3}{8}\biggr\{\frac{\partial^2\Psi}{\partial z^2} + \frac{1}{z}\frac{\partial\Psi}{\partial z}\biggl\} - \frac{\epsilon}{2z^2}\frac{\partial^2\Psi}{\partial\phi^2} + V(\phi)\Psi = i \frac{\partial \Psi}{\partial t}.
 \end{eqnarray}
 This expression is in agreement with Ref. \cite{unruh}.

 In the quantization of the GR Lagrangian in mini-superspace, the time variable is absent due to the time reparametrization freedom. A common approach to recover a time coordinate
 is by introducing matter fields \cite{ruba,demaret,nivaldo1} and using, for example, the Schutz approach to describe the matter content \cite{schutz1,schutz2}. A general expression when the equation of state $p = \omega\rho$, with $\omega$ constant, is used has been obtained in Ref. \cite{nivaldo2}.
Equation (\ref{ea}), without a scalar field, is the same as obtained in Ref. \cite{nivaldo2} (without the ordering factor, which can be directly introduced) for the case corresponding to a cosmological constant, corroborating that Unimodular Gravity has a {\it hidden} connection with General Relativity with a cosmological term.

 \section{Self-adjointness and positivity of energy}

 Is the Hamiltonian operator self-adjoint? What are the conditions for the energy of the system to be bounded from below? These questions are surely relevant since they refer to the problem of consistence and stability of a quantum cosmological approach in Unimodular Gravity, specially in its original formulation. %approach.

 The Hamiltonian operator can be written as,
 \begin{eqnarray}
 \label{h1}
 {\cal H} = \frac{3}{8}\frac{1}{z^2}\frac{\partial  }{\partial z}\biggr(z \frac{ \partial }{\partial z}\biggl) - \frac{\epsilon}{2z^2}\frac{\partial^2}{\partial\phi^2} + V(\phi).
 \end{eqnarray}
 This operator is hermitian if the inner product is given by,
 \begin{eqnarray}
 \label{ip}
 (\phi,\psi) = \int\phi^*\psi zdzd\phi.
 \end{eqnarray}
 This is consistent with the polar form of the Hamiltonian due to the factor ordering that has been introduced. The measure in the internal product (\ref{ip}) has been chosen in order to set the Hamiltonian operator to be Hermitian.

 To verify whether it is self-adjoint, a practical way is to compute the von Neumann deficiency index, which is determined by the number of squared integrable solutions of the equation,
 \begin{eqnarray}
 \label{di}
 {\cal H}\psi = \pm i\psi,
 \end{eqnarray}
 given by $n_\pm$. If $n_+ = n_- = 0$, the operator is self-adjoint, while for $n_+ = n_- \neq 0$, it is not self-adjoint but it admit a self-adjoint extension by choosing a convenient boundary condition. If $n_+ \neq n_-$, the operator is not self-adjoint neither admits self-adjoint extension \cite{rs}.
The solutions of (\ref{di}), with $V(\phi) = 0$ in (\ref{h1}), are given by,
\begin{eqnarray}
\label{sdi}
 \psi_\pm = c_\pm H_\nu^{(1,2)}(x_\pm)e^{-ik\phi},
 \end{eqnarray}
 where $H_\nu^{(1,2)}(x_\pm)$ are Hankel's functions of first and second species, $\nu = \sqrt{-\epsilon}|k|$ and $x_\pm = e^{\mp i\frac{\pi}{4}z}$.
 This solution is divergent under the integration on $\phi$ due to the plane-wave solution resulting from the separation variable method, but this feature can be coped with in the same lines as for
 plane wave solutions  in ordinary quantum mechanics, by interpreting them as distribution. %LF: nao entendi essa frase
 One of the Hankel functions in (\ref{sdi}) presents a divergence for each choice of the sign in (\ref{di}). Hence, the von Neumann deficiency index read $n_+ = n_- = 1$: the Hamiltonian is not self adjoint but admits a self adjoint extension. The issue of the self-adjoint character of the Unimodular Gravity in mini-superespace has been addressed explicitly in Ref. \cite{sorkin} for the positive spatial curvature case, which can be directly applied for the flat curvature case treated here.

The energy associated to the quantum configuration must be bounded from below in order to assure the stability of the system. Essentially, this amounts to verify if the energy can be made positive in order to have a bound from below. Of course, such bound from below is not possible if the energy can be arbitrarily negative: if the energy is bounded from below, a rigid time translation can make it positive, which is impossible if there is no lower limit. %%%LF: achei um pouco estranha esses comentários, repetitivo. Talvez algo como
%%% "The energy associated to the quantum configuration must be bounded from below in order to assure the stability of the system. If the energy is bounded from below, a rigid time translation can make it positive, a necessary condition for stability. If there is no lower limit, i.e. no bound from below, the energy can be arbitratily negative, which leads to an instability."
Hence, the sign of the stationary states with a given energy $E$ must be verified. Stationary states are given by,
 \begin{eqnarray}
 \Psi = \chi e^{iEt},
 \end{eqnarray}
 leading to,
\begin{eqnarray}
 \frac{3}{8}\biggr\{\frac{\partial^2\chi}{\partial z^2} + \frac{1}{z}\frac{\partial\chi}{\partial z}\biggl\} - \frac{\epsilon}{2z^2}\frac{\partial^2\chi}{\partial\phi^2} + V(\phi)\chi = - E\chi.
 \end{eqnarray}

 In order to verify the sign of $E$, the equation (\ref{ea}) is multiplied by $z\chi*$ and integrated on $z$ and $\phi$:
 \begin{eqnarray}
 \int \biggr\{- \frac{3}{8}\frac{\partial \chi^*}{\partial z}\frac{\partial\chi}{\partial z} + \frac{\epsilon}{2z^2}\frac{\partial \chi^*}{\partial \phi}\frac{\partial\chi}{\partial\phi} + V(\phi)\chi^*\chi\biggl\}zdzd\phi = - E\int \chi^*\chi zdzd\phi.
 \end{eqnarray}
 Only if $\epsilon = - 1$ and $V(\phi) \leq 0$ it is assured that $E$ is positive, otherwise, $E$ can assume arbitrary negative values, and the system is unstable by radiative emission.

 For $E > 0$, it is possible to make the redefinitions,,
 \begin{eqnarray}
 z = \sqrt{\frac{3}{8E}}y, \quad \sqrt{\frac{3}{4}}\phi \rightarrow \phi, \quad U(\phi) = \frac{V}{E}.
 \end{eqnarray}
 The Schr\"odinger equation becomes,
 \begin{eqnarray}
 \biggr\{\frac{\partial^2\chi}{\partial y^2} + \frac{1}{y}\frac{\partial\chi}{\partial y}\biggl\} - \frac{\epsilon}{y^2}\frac{\partial^2\chi}{\partial\phi^2} + U(\phi)\chi = - \chi.
 \end{eqnarray}
 Even if the redefinitions above are strictly justified for $\epsilon = - 1$, in the general case we keep the possibility of both signs for $\epsilon$ for the reasons to be discussed below.
 The next step is to determine solutions of this equation. Analytical expressions can be obtained if the potential vanishes.

 \section{Explicit solutions for the wave equation: the case $V(\phi) = 0$}

Fixing $V(\phi) = 0$, the Schrödinger-type equation becomes,
\begin{eqnarray}
 \biggr\{\frac{\partial^2\chi}{\partial y^2} + \frac{1}{y}\frac{\partial\chi}{\partial y}\biggl\} + \chi - \frac{\epsilon}{y^2}\frac{\partial^2\chi}{\partial\phi^2} = 0.
 \end{eqnarray}
 The cases $\epsilon = 1$ and $\epsilon = -1$ will be considered separately.

\subsection{The case $\epsilon = 1$}

For an ordinary scalar field ($\epsilon = 1$), the energy of the system is not positive defined, unless there is no dependence of the wave function on the scalar field, resulting in the same equation considered in Ref. \cite{nivaldo2} but with a different factor ordering. If there is no dependence on the scalar field, implying $E \geq 0$, the dynamic equation becomes,
\begin{eqnarray}
 \biggr\{\frac{\partial^2\chi}{\partial y^2} + \frac{1}{y}\frac{\partial\chi}{\partial y}\biggl\} + \chi  = 0,
 \end{eqnarray}
 with solutions in terms of the original wave function as,
 \begin{eqnarray}
 \psi =  [c_1 J_0(y) + c_2Y_0(y)]e^{iEt},
 \end{eqnarray}
$J_0$ and $Y_0$ being the Bessel and Neumann functions of order $0$. If the wavefunction and/or its first derivative must be made equal 0 at the origin $y = 0$, since it is not defined for $y < 0$, the Neumann function must be discarded, and the solution becomes,
\begin{eqnarray}
 \psi =  cJ_0(y)e^{iEt}.
 \end{eqnarray}

\subsection{The case $\epsilon = - 1$}

For a phantom scalar field, for which $E \geq 0$, the equation becomes,
\begin{eqnarray}
 \biggr\{\frac{\partial^2\chi}{\partial y^2} + \frac{1}{y}\frac{\partial\chi}{\partial y}\biggl\} + \chi +  \frac{1}{y^2}\frac{\partial^2\chi}{\partial\phi^2} = 0
 \end{eqnarray}
 In order to have finite wave functions, the condition
 \begin{eqnarray}
\frac{\partial^2\chi}{\partial\phi^2} = - k^2,
\end{eqnarray}
must be imposed, leading to
\begin{eqnarray}
\chi(y,\phi) = \chi(y)e^{\pm ik\phi}.
\end{eqnarray}

The Schr\"odinger like equation reduces to,
\begin{eqnarray}
 \frac{\partial^2\chi}{\partial y^2} + \frac{1}{y}\frac{\partial\chi}{\partial y} + \biggr\{1 - \frac{k^2}{y^2}\biggl\}\chi = 0.
 \end{eqnarray}
 The solutions (returning to the original variables) are,
 \begin{eqnarray}
 \psi(z,\phi,t) = \biggr\{A_1 J_{|k|}\biggr(\sqrt{\frac{8E}{3}}z\biggl) + A_2 J_{-|k|}\biggr(\sqrt{\frac{8E}{3}}z\biggl)\biggl\}e^{i \left(Et \pm \sqrt{\frac{3}{4}}k\phi\right)},
 \end{eqnarray}
 with $A_{1,2}$ being integration constants.

 Choosing $A_1 = A_2 = A$, the solution can be written in a unified way:
 \begin{eqnarray}
 \label{e-}
 \psi(z,\phi,t) = A J_k\biggr(\sqrt{\frac{8E}{3}}z\biggl) e^{i \left(Et \pm \sqrt{\frac{3}{4}}k\phi\right)},
 \end{eqnarray}
 with $ - \infty < k < + \infty$.

 \section{Constructing the wave packet}

 We proceed now by constructing the wave packet.  In a general form it is given by,
 \begin{eqnarray}
 \Psi = \int A(k,E) J_\nu(\beta \sqrt{E}) e^{-iEt \pm ik\phi}dE dk.
 \end{eqnarray}
 It is more convenient to define $x = \sqrt{E}$. The integral can be rewritten as,
 \begin{eqnarray}
 \Psi = \int \bar A(k,x) J_\nu(\beta x) e^{-ix^2t \pm ik\phi}dx dk,
 \end{eqnarray}
 with $\bar A(k,x) = 2x A(k,x)$ and $\beta = \sqrt{8/3}z$.

 An explicit wave packet expression can be obtained by choosing $\bar A(k,x)$ such that,
 \begin{eqnarray}
 \label{wp}
 \bar A = \bar A_0e^{-\lambda k^2} x^{\nu + 1}e^{- \gamma x^2},
 \end{eqnarray}
 where $\gamma$ and $\lambda$ are positive real numbers, and formulas ($6.631-6.634$) of Ref. \cite{gr} were applied, the same used in Refs. \cite{nivaldo2,brasil}.

 Again, the cases $\epsilon = \pm 1$ will be analyzed separately.

 \subsection{The case $\epsilon = 1$}

 In this case, $\nu = 0$, and the wave packet reads
 \begin{eqnarray}
 \psi(z,t) = \psi_0\frac{e^\frac{- z^2}{4\alpha}}{\alpha},
 \end{eqnarray}
 where $\alpha = \gamma + it$ and $\psi_0$ is a normalization constant factor.
 The expectation value of the scale factor is given by,
 \begin{eqnarray}
 <a> &=& \int_0^\infty \psi^* a \psi zdz = \int_0^*\psi^* z^\frac{4}{3}\psi dz,\nonumber\\
&=& \gamma^2\int_0^\infty \frac{e^{- \gamma z^2}{B}}{B}z^\frac{4}{3} dz,
\end{eqnarray}
where $B = \gamma^2 + t^2$. Defining $x = z/\sqrt{B}$ and performing the integration, the expectation value for the scale factor becomes,
\begin{eqnarray}
<a> \propto (\gamma^2 + t^2)^{1/6},
\end{eqnarray}
describing a symmetrical bounce. For $t \rightarrow \infty$, the expectation value approaches $<a> \propto t^{1/3}$. This corresponds to the usual exponential de Sitter phase written in terms of time coordinate with $N = a^{-3}$: the Unimodular quantum model has a "hidden" cosmological constant, as it happens in its classical version.

If $k \neq 0$ would have been allowed, solutions with $E < 0$ should be included, and this would imply solutions for the wavefunction in terms of modified Bessel functions, which have an exponential behaviour asymptotically, reflecting the instability due to the unbounded energy spectra.

\subsection{The case $\epsilon = - 1$}

For a phantom scalar field characterized by $\epsilon = - 1$, the scalar field $\phi$ may be taken into account without spoiling the positiveness of $E$. Hence, the parameter $k$ is present in the wave function. The solution is given by (\ref{e-}). The superposition can be constructed by using the factor (\ref{wp}).

 The wave packet reads, after the integrations in $E$ and $k$:
 \begin{eqnarray}
 \Psi = \Psi_0\frac{e^\frac{-\beta}{4\alpha}}{\alpha}\exp\left[\frac{\ln^2\left(\frac{\beta}{2\alpha}\right)}{4\lambda^2} \pm i\ln\biggr(\frac{\beta}{2\alpha}\biggl)\frac{\phi}{2\lambda} - \frac{\phi^2}{4\lambda}\right].
 \end{eqnarray}
 The expectation value for the scale factor and for the scalar field are given by,
 \begin{eqnarray}
 <a> &\propto& (\gamma^2 + t^2)^{1/6},\\
 <\phi> &\propto& \mbox{constant}.
 \end{eqnarray}
 Again, the expectation value for the scale factor coincides with those found for $\epsilon = + 1$ and with the results obtained in Ref. \cite{nivaldo2}, in which the time is fixed by a fluid representing the vacuum energy. Even if the scalar factor is allowed to be dynamical, the expectation value indicates a frozen scalar field. This is in agreement with the results of Ref. \cite{brasil} in the context of scalar-tensor theories of gravity with the time coordinate given by the matter variables.

 \section{The quantum unimodular scalar-tensor gravity}

 In order to discuss the scalar-tensor theory case, we will be consider the Brans-Dicke (BD) theory \cite{bd}, which is given by the action,
 \begin{eqnarray}
 \label{bd1}
 {\cal A} = \int d^4x\biggr(\phi R - \omega\frac{\phi_{;\rho}\phi^{;\rho}}{\phi}\biggl).
 \end{eqnarray}
 In this expression, $\phi$ is a scalar field whose inverse is connected to the (dynamical) gravitational coupling $\phi$, and $\omega$ is a dimensionless parameter. In general, the General Relativity limit is achieved by imposing $\phi = $ constant and $\omega \rightarrow \infty$. The Unimodular version of the Brans-Dicke theory has been discussed in Ref. \cite{fabris}.

 The BD theory is originally formulated in the non-minimal coupling, but it can be recast in the minimal coupling by making the transformation $g_{\mu\nu} = \phi^{-1}\tilde g_{\mu\nu}$, resulting in the action,
 \begin{eqnarray}
 \label{bd2}
 {\cal A} = \int d^4x\biggr(\tilde R - \tilde\omega\frac{\phi_{;\rho}\phi^{;\rho}}{\phi^2}\biggl),
 \end{eqnarray}
  with $\tilde\omega = \omega + 3/2$. In this frame, it is easy to identify when the energy of the scalar field is positive ($\omega > - 3/2$) or negative ($\omega < - 3/2$).
  The equation resulting from the quantization of the action (\ref{bd1}) and transformed action (\ref{bd2}) are also connected by the same conformal transformation \cite{colistete}. However, the specific prediction for the behavior, for example, of the scale factor, may depend on which frame it is considered.

 An analysis of quantum models resulting from (\ref{bd2}) has been done in Refs. \cite{brasil,colistete}. In particular, in Ref. \cite{brasil} extra fields motivated by string theories have been introduced together with a fluid component described through the Schutz formalism, allowing to recover a notion of time which is associated with the fluid degrees of freedom. The expectation value of the scalar field indicates no dynamics: the model reduces to the GR, at least in what concerns the expectation values.
  In what follows the time is recovered through the imposition of the unimodular condition as described in the previous section.

  The Lagrangian associated with the action (\ref{bd1}), after introducing the flat FLRW metric and performing an integration by parts, reads,
  \begin{eqnarray}
  {\cal L} = 6\frac{\phi a\dot a^2}{N} + 6\frac{\dot\phi a^2 \dot a}{N} - \frac{\omega}{N}a^2\frac{\dot\phi^2}{\phi}.
  \end{eqnarray}
  Imposing the Unimodular condition $N = a^{-3}$ and performing the conformal transformation, the Lagrangian becomes,
  \begin{eqnarray}
  {\cal L} = 6b^4\dot b^2 - \tilde\omega b^6\frac{\dot\phi^2}{\phi^2}.
  \end{eqnarray}
  Defining $\epsilon = \frac{\tilde \omega}{|\tilde\omega|} = \pm 1$ and
  \begin{eqnarray}
  \sigma = \sqrt{|\tilde\omega|}\ln \phi,
  \end{eqnarray}
  the Lagrangian takes the form,
  \begin{eqnarray}
  {\cal L} = \biggr(6b^4\dot b^2 - \epsilon b^6\dot\sigma^2\biggl)e^{-2\sigma/\sqrt{|\tilde\omega|}}.
  \end{eqnarray}

  The Legendre transformation leads to the Hamiltonian,
  \begin{eqnarray}
  {\cal H} = e^{2\sigma/\sqrt{|\tilde \omega}}\biggr\{\frac{1}{24}\frac{\pi^2_b}{b^4} - \frac{\epsilon}{4}\frac{\pi_\sigma^2}{b^6}\biggl\}.
  \end{eqnarray}
  The differential equation for the stationary state of energy $E$ is
  \begin{eqnarray}
  \label{st1}
 - \frac{1}{24}\frac{\partial^2_b\Psi}{b^4} + \frac{\epsilon}{4}\frac{\partial_\sigma^2\Psi}{b^6} = Ee^{-2\sigma/\sqrt{|\tilde \omega}}\Psi.
  \end{eqnarray}
  An important remark here is that equation (\ref{st1}) does not admit the separation of variables method. In principle this brings the necessity of using, for example, numerical methods. This is in opposition to what happens in the original Brans-Dicke case with a fluid described by Schutz formalism, for which the conformal transformation decouples the equations, which can then be solved using the separation of variables method. 

  It must also be remarked that the computation has been done by imposing the Unimodular condition $N = a^{-3}$ before performing the conformal transformation. If the order is reversed, and the conformal transformation is made before imposing the unimodular condition, a new lapse function $\tilde N$ appears such that $N = \phi^{-1/2}\tilde N$. If the Unimodular condition is applied for the transformed determinant of the metric, $\tilde g$, the resulting Lagrangian is,
  \begin{eqnarray}
  {\cal L} = 6b^4\dot b^2 - \epsilon b^6\dot\sigma^2.
  \end{eqnarray}
  This Lagrangian is the same as the Lagrangian in GR (\ref{lug}), with the identifications $a \rightarrow b$ and $\phi \rightarrow \sqrt{\sigma}$. Hence, all the results found before apply here.
   To be more explicit, for $\epsilon = + 1$, in order to have a bounded energy spectrum, the wave function must not depend on the separation parameter $k$. That amounts to having a constant $\phi$, and the conformal transformation is trivial -- just a multiplication by a constant. For $\epsilon = - 1$, the scalar field becomes dynamical, but the conformal transformation amounts to a simple redefinition of the separation parameter $k$. In this sense, all computed expectation values remain the same, including for the scale factor.

  To impose the Unimodular condition after the conformal transformation corresponds to another structure since the determinant of the metric in this different frame is connected by $g = Na^3 = \phi^{-2}\tilde N b^3 = \phi^{-2}\tilde g$. Evidently, to impose $g = 1$ is a
  different condition with respect to imposing $\tilde g = 1$.

  \section{Conclusions}

 The problem of the construction of a quantum cosmological model in the minisuperspace in the framework of Unimodular Gravity has been discussed. The pioneer work of \cite{unruh} has been revisited. In this reference, the QC model has been implemented using the original UG constraint $\sqrt{-g} = 1$. This condition implies to fix the lapse function, breaking the invariance with respect to time reparametrization of the QC formulation in GR. Hence, the Hamiltonian constraint is no longer valid, and a time coordinate appears driven by the Hamiltonian as in the usual Quantum Mechanics. In Ref. \cite{unruh} it has been shown that the UG QC leads to a de Sitter phase asymptotically. Here, we have extended the analysis of Ref. \cite{unruh} to a phantom scalar field (in that work only an ordinary scalar field had been considered), and we argue that only in the phantom case a non-trivial behavior of the scalar field can be find: in the ordinary case, the scalar field must be constant in order to have an energy bounded from below.
 \\
 We succeeded in constructing a wave packet, obtaining explicit expressions for the expectation values of the scale factor and the scalar field (when it is not trivially constant).  For the scale factor, the results coincide with the case of a cosmological constant described through the Schutz formalism with the explicit solutions set down in Refs. \cite{nivaldo1,nivaldo2}. Hence, as in the classical UG framework, in its quantum version the cosmological constant is {\it hidden} in the UG structure.
 \\
 We then extended the analysis to scalar-tensor theories, in the case of the Brans-Dicke one. To solve the equations is necessary to perform a conformal transformation, rephrasing the theory in its minimal coupled version (the Einstein frame). The final formulation depends if the unimodular condition is imposed before or after the conformal transformation. Only in the second case it is possible to obtain an analytical solution, which leads essentially to the same scenario as in GR: a constant expectation value for the scalar field, and a behavior for the (transformed) scale factor identical to that found in GR framework with a cosmological constant described by the Schutz variable.
 \\
 A final remark concerns the unimodular constraint. If instead of $\sqrt{-g} = 1$ this constraint is implemented through a fiducial external tensorial density $\chi$ such that $\sqrt{-g} = \chi$, the lapse function becomes again an arbitrary Lagrange multiplier connected with the Hamiltonian which, as a consequence, must vanish: the usual problem of time of QC in GR context is recovered.

 \bigskip
 \noindent
 {\bf Acknowledgement:} NPN acknowledges the support of CNPq of Brazil under grant PQ-IB 310121/2021-3. We thank CNPq (Brazil), FAPES (Brazil) and Fundação Araucária (Brazil) for partial financial support.

\end{document}